\documentclass[aps,
twocolumn,
showpacs,prl]{revtex4}
\usepackage{epsf}
\begin{document}
\title{Microcanonical Analyses of Peptide Aggregation Processes}
\author{Christoph Junghans}
\email[E-mail: ]{Christoph.Junghans@itp.uni-leipzig.de}
\author{Michael Bachmann}
\email[E-mail: ]{Michael.Bachmann@itp.uni-leipzig.de}
\author{Wolfhard Janke}
\email[E-mail: ]{Wolfhard.Janke@itp.uni-leipzig.de}
\homepage[\\ Homepage: ]{http://www.physik.uni-leipzig.de/CQT}
\affiliation{Institut f\"ur Theoretische Physik and Centre for Theoretical Sciences (NTZ), 
Universit\"at Leipzig, Augustusplatz 10/11, D-04109 Leipzig, Germany}
\begin{abstract}
We propose the use of microcanonical analyses for numerical studies of peptide aggregation
transitions.
Performing multicanonical Monte Carlo simulations 
of a simple hydrophobic-polar continuum model for interacting heteropolymers
of finite length, we find that
the microcanonical entropy behaves convex in the transition region, leading to a 
negative microcanonical specific heat.
As this effect 
is also seen in first-order-like transitions of other finite systems, our
results provide clear evidence for recent hints that the characterisation of phase separation
in first-order-like transitions of finite systems profits from this microcanonical view. 
\end{abstract}
\maketitle
Thermodynamic phase transitions in macroscopic, infinitely large systems are typically 
analysed in the thermodynamic limit 
of a canonical ensemble, i.e., the temperature $T$ is treated as an intensive external 
control parameter adjusted by the heat bath, and the total system energy $E$ is distributed
according to the Boltzmann-Gibbs statistics. The probability for a macrostate with energy $E$ is
given by $p(E)=g(E)\exp(-E/k_BT)/Z$, where $g(E)$ is the density of states, $Z$ the partition sum,
and $k_B$ the Boltzmann constant. As long as the microcanonical entropy $S(E)=k_B\ln g(E)$ is
a concave function of $E$, the microcanonical (caloric) temperature
$T(E)=(\partial S(E)/\partial E)^{-1}$ 
for fixed volume $V$ and particle number ${\cal N}$ is a monotonically increasing function of $E$. 
Consequently, the microcanonical specific heat 
$C_V(E)=\partial E/\partial T(E)=-(\partial S/\partial E)^2/(\partial^2 S/\partial E^2)$
is positive. The specific heat can only become negative in an energetic regime, where $S(E)$
is convex. In this region, the caloric $T(E)$ curve exhibits a typical backbending, which means 
that the system
becomes colder with increasing total energy. For this reason, the temperature $T$ is not the
most appropriate control parameter and the analysis of such, in particular finite, systems is
more adequately performed in the microcanonical ensemble, where the system energy $E$
is considered as the adjustable external parameter~\cite{gross1,gross2}.

It is a surprising fact that the backbending effect is indeed observed in transitions
with phase separation. Although this phenomenon has already been known for a long time 
from astrophysical systems~\cite{thirring1}, it has been widely ignored since then 
as somehow ``exotic'' effect.
Recently, however, experimental evidence was found from melting studies of sodium clusters
by photofragmentation~\cite{schmidt1}. Bimodality and negative specific heats are also
known from nuclei fragmentation experiments and models~\cite{pichon1,lopez1}, as well as 
from spin models on finite lattices which experience first-order transitions in the thermodynamic
limit~\cite{wj1,pleimling1}. This phenomenon is also observed in a large number of 
other isolated finite model systems for evaporation and melting 
effects~\cite{wales1,hilbert1}. 

In this Letter, we demonstrate the usefulness of the microcanonical ensemble for studies of
the aggregation process of small proteins (peptides), which, due
to the fixed inhomogeneous sequence of amino acids, are necessarily systems
of finite size. Understanding protein 
aggregation is essential not only for gaining insights into general mechanisms of protein
folding, but also for unraveling the reasons of human diseases caused by protein clustering.
A well-known example is associated with Alzheimer's disease, where a few identical small fragments
of large proteins show the tendency to form fibrils, e.g., the
hydrophobic A$\beta_{16-22}$ segment of the $\beta$-amyloid peptide
A$\beta$~\cite{irb1}.

Our results are based on computer simulations of a simple continuum aggregation model for heteropolymers.
Since the hydrophobic force governs the tertiary folding process resulting 
in a compact hydrophobic core surrounded by a shell of mainly polar residues, in our model 
the 20 amino acids naturally occuring in proteins are classified as hydrophobic ($A$) and 
polar ($B$)~\cite{dill1}. For the individual peptides, we employ the AB model~\cite{still1} 
in three spatial dimensions. At a mesoscopic length scale, this coarse-grained model 
with virtual peptide bonds and virtual bond angles has proven quite successful in the 
qualitative characterisation of naturally observed protein folding channels~\cite{ssbj1}.  
Keeping the same parameter sets for the interaction of monomers of {\em different} polymers,
the model for the aggregate reads:
\begin{equation}
\label{eq:aggmod}
E=\sum\limits_{\mu} E_{\rm AB}^{(\mu)}+\sum\limits_{\mu<\nu} 
\sum_{i_\mu,j_\nu}\Phi(r_{i_\mu j_\nu};\sigma_{i_\mu},\sigma_{j_\nu}),
\end{equation}
where $\mu,\nu$ label the $M$ polymers interacting with each other, and 
$i_\mu,j_\mu$ index the $N$ monomers of the $\mu$th polymer whose
intrinsic energy is given by
\begin{equation}
\label{eq:abmod}
E_{\rm AB}^{(\mu)}=\frac{1}{4}\sum\limits_{i_\mu=1}^{N-2}(1-\cos \vartheta_{i_\mu})+%
\!\!\sum\limits_{j_\mu>i_\mu+1}\Phi(r_{i_\mu j_\mu};\sigma_{i_\mu},\sigma_{j_\mu}),
\end{equation}
with $0\le \vartheta_{i_\mu}\le \pi$ denoting the bending angle between monomers 
$i_\mu$, $i_\mu+1$, and $i_\mu+2$.
The nonbonded inter-residue pair potential 
\begin{equation}
\label{eq:phi}
\Phi(r_{i_\mu j_\nu};\sigma_{i_\mu},\sigma_{j_\nu})=
4\left[r_{i_\mu j_\nu}^{-12}-C(\sigma_{i_\mu},\sigma_{j_\nu})r_{i_\mu j_\nu}^{-6}\right]
\end{equation}
depends on the distance $r_{i_\mu j_\nu}$ between the residues, and on their type,
$\sigma_{i_\mu}=A,B$. The long-range behavior is attractive for 
like pairs of residues [$C(A,A)=1$, $C(B,B)=0.5$] and repulsive else [$C(A,B)=C(B,A)=-0.5$]. 
The lengths of all virtual peptide bonds are set to unity. 

In our aggregation study, we have performed multicanonical simulations~\cite{muca1} 
for two identical peptides 
with 13 monomers and the sequence 13.1: $AB_2AB_2ABAB_2AB$ is arbitrarily chosen from the Fibonacci 
series~\cite{still1}. For consistency, the simulations were repeated for pairs of 
identical homopolymers, 
$2\times A_{13}$ and $2\times B_{13}$, as well as the larger aggregates $3\times 13.1$ and 
$4\times 13.1$~\cite{jbj1}. In all cases, aggregation behaviors of similar type 
as for the two-peptide system were identified. 

For the simulations of the $2\times 13.1$ system, the peptides were
confined in a periodic cube with edge lengths $L=40$. 
We varied the edge lengths to make sure that effects due to this confinement are negligible.
A sequence of spherical-cap 
updates~\cite{baj1} and three-monomer corner 
rotations ensured an ergodic scan of the conformational space. 
After performing 180 
multicanonical recursions, a total number of $2\times 10^{10}$ updates was generated.
The primary result of these simulations is, up to
an unimportant constant, the
density of states $g(E)$ which has been precisely estimated over about 100 orders
of magnitude. 

\begin{figure}
\centerline{\epsfxsize=8.0cm \epsfbox{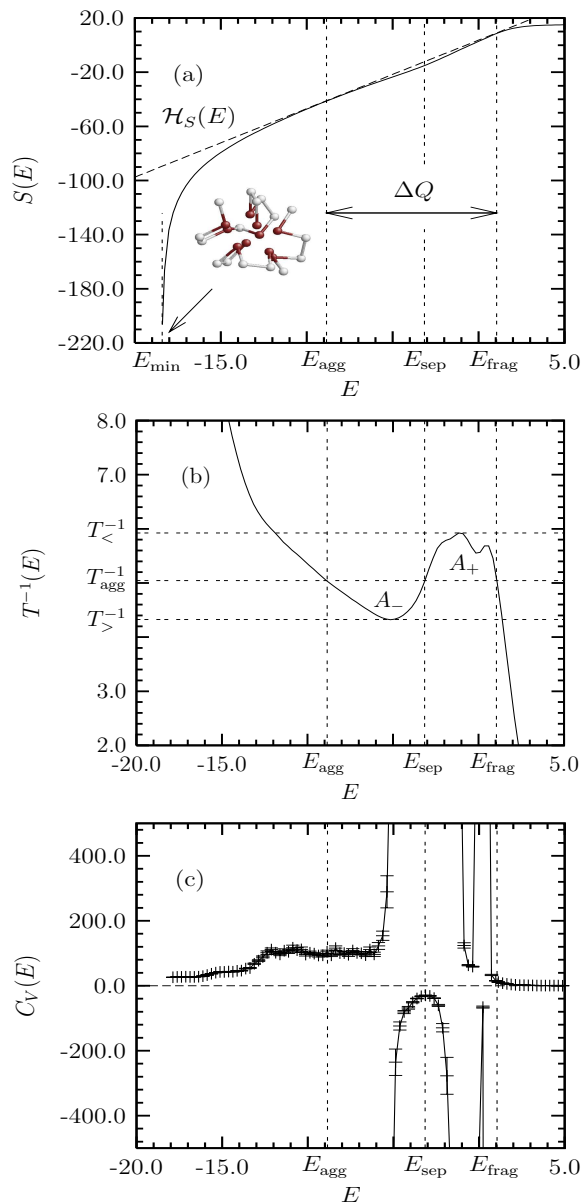}}
\caption{\label{fig:micro} 
Aggregation transition from the microcanonical perspective: (a) microcanonical entropy $S(E)$
(up to a constant) and concave hull ${\cal H}_S(E)$,
(b) inverse caloric temperature $T^{-1}(E)$, 
and (c) specific heat $C_V(E)$. The errors are very small and therefore only shown for $C_V(E)$.
}
\vspace*{-5mm}
\end{figure}

In Fig.~\ref{fig:micro}(a), the microcanonical entropy $S(E)= \ln g(E)$ ($k_B\equiv 1$)~\cite{rem1}
is plotted (up to an unimportant additive 
constant) for the two-peptide system, ranging from the aggregate phase including 
the lowest energy
found in the simulation ($E_{\rm min} =E_{\rm AB,min}^{(1)}+E_{\rm AB,min}^{(2)}+
E_{\rm AB,min}^{(1,2)} \approx -18.407$), to the phase of the fragmented
polymers. The conformation of the lowest-energy aggregate
has a two-cap-like, globular shape with a compact hydrophobic
core jointly formed by the two heteropolymers, see the inset of Fig.~\ref{fig:micro}(a). 
It should be noted that the 
individual conformations in the aggregate strongly differ from the single-peptide
ground states ($E_{\rm min}^{\rm single}\approx -4.967$~\cite{baj1}) and 
their respective energies in the aggregate are much larger ($E_{\rm AB,min}^{(1)}\approx -3.197$,
$E_{\rm AB,min}^{(2)}\approx -3.798$). The strongest contribution is due to the interaction
between the heteropolymers ($E_{\rm AB,min}^{(1,2)}\approx -11.412$). 

The most interesting region in Fig.~\ref{fig:micro}(a) is the phase coexistence regime
$E_{\rm agg}\approx -8.85 \le E\le 1.05\approx E_{\rm frag}$, where the entropy 
exhibits a convex intruder.  The concave hull 
${\cal H}_S(E)=S(E_{\rm agg})+E/T_{\rm agg}$, which is the
tangent connecting $S(E_{\rm agg})$
and $S(E_{\rm frag})$, is the Gibbs construction. Its slope defines the 
inverse of the aggregation temperature $T_{\rm agg}\approx 0.198$. The interval 
$\Delta Q=E_{\rm frag}-E_{\rm agg}=T_{\rm agg} [S(E_{\rm frag})-S(E_{\rm agg})]\approx 9.90$ 
is the latent heat required 
to release inter-chain contacts at the aggregation temperature $T_{\rm agg}$.
The energy, where the difference $\Delta S(E)={\cal H}_S(E)-S(E)$ is maximal, is denoted as
$E_{\rm sep}$ and the associated maximum deviation is the surface entropy
$\Delta S_{\rm surf}\equiv \Delta S(E_{\rm sep})$. 
The derivative of the Gibbs construction
gives the Maxwell line $T_{\rm agg}^{-1}={\rm const}\approx 5.043$ in the 
reciprocal caloric $T^{-1}(E)$ curve which is shown in 
Fig.~\ref{fig:micro}(b). A bijective mapping between $T$ and $E$ is 
only possible for $T>T_>\approx 0.231$ and $T<T_<\approx 0.169$. 
This means, for values above $T_>$ and below $T_<$, that
the temperature $T$ is a useful control parameter. The two-heteropolymer system 
forms an aggregate for $T<T_<$, where the separation into individual polymers is
not useful because inter-polymer attraction dominates over intrinsic structure formation
and the aggregate determines the mesoscopic length and energy scale. On the other hand, 
for $T>T_>$ the polymers are only weakly interacting fragments, i.e., they can be considered
separately, the total system energy is an \emph{extensive} variable 
($E\approx E_{\rm AB}^{(1)}+ E_{\rm AB}^{(2)}$). 
\emph{Only in these two temperature regions}, the interpretation of
the canonical formalism is generic.

In the transition region $T_<\le T\le T_>$, however, the interaction strength between
the polymers is as strong as intrinsic monomer-monomer attraction and cannot be neglected. 
As a consequence of the convexity of $S(E)$ in the
interval $E_{\rm agg}<E<E_{\rm frag}$, there is no one-to-one correspondence 
between temperature and energy in the transition regime
which results in the backbending effect: Fragmentation of the aggregate leads to
a decrease of temperature, although the system energy increases. The areas
$A_+=T^{-1}_{\rm agg}(E_{\rm sep}-E_{\rm frag})-[S(E_{\rm frag})-S(E_{\rm sep})]$ 
and $A_-=T^{-1}_{\rm agg}(E_{\rm sep}-E_{\rm agg})-[S(E_{\rm sep})-S(E_{\rm agg})]$ 
formed by the Maxwell line and the $T^{-1}(E)$ curve as shown in Fig.~\ref{fig:micro}(b) 
are identical. These areas
determine the interfacial entropy $\Delta S_{\rm surf}=A_+=A_-$~\cite{wj1}, which
is interpreted as the loss of entropy due to the existence of the phase 
boundary~\cite{gross3} between the aggregate and the fragment macrostates of the
polymers. Consequently, as the energy of the total system is 
not extensive in the transition region, $E$ is the favored control parameter compared with 
$T$. Therefore, the aggregation transition is more favorably analysed in the
microcanonical ensemble, at least for such finite systems like the heteropolymers in our
study, where an extension towards the thermodynamic limit is not possible.

The most remarkable result is the negativity of the specific heat of the system in the
backbending region, as shown in Fig.~\ref{fig:micro}(c).
A negative specific heat in the phase separation regime is due to the nonextensitivity
of the energy of the two subsystems resulting from the interaction between the 
polymers. ``Heating'' a \emph{large} aggregate would lead to the stretching of 
monomer-monomer contact distances, i.e., the potential energy of an exemplified pair
of monomers increases, while kinetic energy and, therefore, temperature remain 
widely constant. In a comparatively \emph{small} aggregate, additional energy leads
to cooperative rearrangements of monomers in the aggregate in order to reduce surface tension, 
i.e, the formation of molten globular aggregates is suppressed.
In consequence, kinetic energy is transfered into potential energy and the temperature
decreases. In this regime, the aggregate becomes colder, although the total energy
increases. 

\begin{figure}
\centerline{\epsfxsize=8.0cm \epsfbox{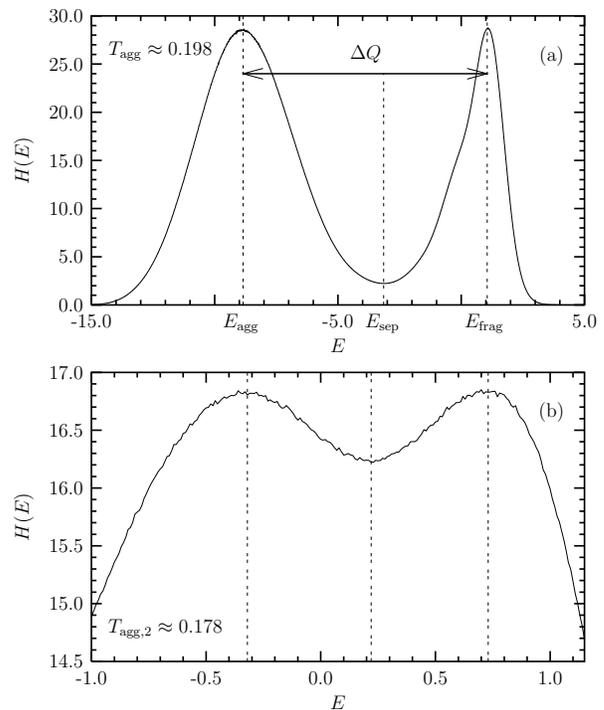}}
\caption{\label{fig:pofe} 
Bimodal canonical energy distribution close to the (a) aggregation temperature
$T_{\rm agg}$ and (b) the subphase transition near $T_{\rm agg,2}$. 
Vertical dashed lines mark extremal points. The corresponding
energies in (a) are identical with those indicating the phase boundaries in 
Fig.~\ref{fig:micro} and, in particular, to the crossing points of the reciprocal caloric
curve $T^{-1}(E)$ with the Maxwell line in Fig.~\ref{fig:micro}(b).
}
\vspace*{-5mm}
\end{figure}
Figure~\ref{fig:pofe}(a) shows the typical bimodal canonical energy distribution
$H(E)\sim g(E)\exp(-E/k_BT_{\rm agg})$ close to the transition temperature
$T_{\rm agg}$. The maximum points are identical with the
energies of the phase boundaries, $E_{\rm agg}$ and $E_{\rm frag}$, and the
minimum is found at $E_{\rm sep}$~\cite{wj1}. For this reason, the 
difference of the energies belonging to the maximum points of the canoncial 
distribution is identical with the latent heat $\Delta Q$. 
The minimum of this distribution coincides with the energy 
$E_{\rm sep}\approx -3.15$, where the Maxwell line
crosses the $T^{-1}(E)$ curve in the backbending regime in Fig.~\ref{fig:micro}(b).
These identifications are easily proven by setting the logarithmic 
derivative of $H(E)$ at $T_{\rm agg}$ to zero, which yields 
$\partial S(E)/\partial E = T^{-1}_{\rm agg}$. The left-hand side is the 
reciprocal microcanonical temperature and thus $T^{-1}(E)=T^{-1}_{\rm agg}$. As is seen from 
Fig.~\ref{fig:micro}(b), this equation has \emph{three} solutions, at
$E_{\rm agg}$, $E_{\rm sep}$, and $E_{\rm frag}$. Therefore, $H(E)$ possesses three extremal
points at exactly these energies. Another expected result of the correspondence between
the canonical and microcanical representations is that the interfacial surface entropy
can be written as: $\Delta S_{\rm surf}=k_B\ln(H(E_{\rm agg})/H(E_{\rm sep}))=%
k_B\ln(H(E_{\rm frag})/H(E_{\rm sep}))$~\cite{wj1}. These expressions serve as convenient 
estimators of the 
surface tension, which can be defined as 
$\sigma=T_{\rm agg}\Delta S_{\rm surf}/R^2_{\rm agg}$, 
where $R^2_{\rm agg}$ is the square radius of gyration of the aggregate.

A short remark shall also be devoted to a second, much weaker transition
that accompanies the aggregation transition. It is also of ``backbending'' type
and can be observed in the fragmentation region in Figs.~\ref{fig:micro}(b) and (c)
close to $E\approx -0.32$. The associated transition temperature is $T_{\rm agg,2}\approx 0.178$
and is, therefore, \emph{smaller} than $T_{\rm agg}$, but happens in the energetic region,
where the population of \emph{fragmented} macrostates dominates. In fact, this effect is 
difficult to understand and requires a system parameter that allows the structural 
discrimination between macrostates. A detailed microcanonical analysis of the square 
relative distance between the centers of masses of the polymers reveals~\cite{jbj1}
that for energies close to $E_{\rm frag}$ the system is in a fragmented state,
and the population of aggregated polymers in this energy region is extremely small. 
The situation is different for energies $E < 0.22$, 
where weakly stable
aggregated conformations and polymer fragments coexist. Only for much smaller 
energies ($E<E_{\rm agg}$), compact aggregates dominate. Having this
in mind, the transition can also be understood from the canonical view.
For temperatures below $T_{\rm agg,2}\approx 0.178$, stable aggregates (solids)
of low energies ($E<E_{\rm agg}$) dominate. Approaching $T_{\rm agg,2}$, the 
system enters the subphase of coexisting unstable pre-molten aggregates of comparatively 
high energies ($E\approx -0.32$) and already fragmented peptides.
From Fig.~\ref{fig:micro}(b) we see that this process 
is also accompanied by \emph{cooling} due to monomer arrangements reducing 
surface tension.
These monomer translocations are, however, energetically unfavorable. Eventually, for temperatures
larger than $T_{\rm agg}$, conformations of weakly coupled separate fragments 
(liquid) dominate. 
The intermediary subphase is never dominating, 
and therefore unstable. 
After these remarks this transition is already visible in the
cusp-like behavior of the left, inner wing of the right fragmentation 
peak in Fig.~\ref{fig:pofe}(a). Reweighting to the subphase transition temperature
$T_{\rm agg,2}$, the bimodal structure of the canonical energy distribution
in this energy range is clearly revealed in Fig.~\ref{fig:pofe}(b).
Compared with the distribution at the aggregation transition in Fig.~\ref{fig:pofe}(a), 
the ratio between maximum and minimum is small and, therefore, also the surface 
tension. In consequence, the transition between the solid and the pre-molten, unstable
aggregates is, compared with the aggregation transition, negligibly weak. 
Note that the aggregation peak, not shown in 
Fig.~\ref{fig:pofe}(b), is much more pronounced than 
the peaks of the pre-molten
aggregates at $E\approx -0.32$ and fragments at $E\approx 0.73$.

In this Letter, we have shown by employing a mesoscopic hydrophobic-polar 
heteropolymer aggregation model that the aggregation transition is a 
phase separation process, where
the loss of entropy due to the existence of the phase boundary results in 
negative specific heat. This is an effect which is guided by 
changes of the interfacial entropy as a result of surface effects. Therefore, 
this effect is expected to disappear in the thermodynamic limit of 
macroscopic systems. It should strongly be emphasized, however, that peptides and 
proteins, like the exemplified model heteropolymers used in our study, are 
\emph{necessarily} systems of \emph{finite} length and a thermodynamic limit cannot be 
defined. For this reason, standard
canonical formalisms for the analysis of conformational pseudophase transitions with 
phase separation are not suitable for these systems, since the temperature is not a unique 
control parameter and the total system energy measured in units of energy scales of 
mesoscopic particles (e.g., aggregates or single polymers) is not an extensive, separable 
quantity. In such cases, microcanonical thermodynamics with the energy itself as the external
control parameter provides a more favorable basis for the study of first-order-like 
transitions.
The interesting phenomenon of the negativity
of the microcanonical specific heat in peptide
aggregation should be motivation for an experimental verification which is still pending.  

We thank Klaus Kroy for helpful discussions. This work is partially supported by 
DFG Grant No.\ JA 483/24-1 and
NIC J\"ulich JUMP/JUBL supercomputer Grant No.\ hlz11.

\begin{thebibliography}{199}
\bibitem{rem1}
The Hertz definition of the entropy is ${\cal S}(E)=k_B\ln \Gamma(E)$, where 
$\Gamma(E)=\int_{E'<E}\, dE'\, g(E')$ is the phase-space volume. In the thermodynamic limit
${\cal S}(E)=S(E)$, but for finite systems, as in our case, these two entropies are not necessarily
identical~\cite{hilbert1}. We have, however, checked our results using both entropy definitions
and did not find noticeable deviations in the transition regimes.
In particular, the transition temperatures are virtually identical within the
numerical precision of our data.
\end{thebibliography}
\end{document}